\documentclass[authorversion]{acmart} 

\AtBeginDocument{%
  }

\copyrightyear{2024}
\acmYear{2024}
\setcopyright{acmlicensed}\acmConference[CHI '24]{Proceedings of the CHI Conference on Human Factors in Computing Systems}{May 11--16, 2024}{Honolulu, HI, USA}
\acmBooktitle{Proceedings of the CHI Conference on Human Factors in Computing Systems (CHI '24), May 11--16, 2024, Honolulu, HI, USA}
\acmDOI{10.1145/3613904.3642855}
\acmISBN{979-8-4007-0330-0/24/05}

\newcommand{\Revision}{\textcolor{black}}

\begin{document}

\title{Machine Learning Processes as Sources of Ambiguity: Insights from AI Art}



\author{Christian Sivertsen}
\email{csiv@itu.dk}
\orcid{0000-0001-6843-5009}
\affiliation{%
  \institution{IT University of Copenhagen}
  \streetaddress{Rued Langgaards Vej 7}
  \city{Copenhagen}
  \country{Denmark}
  \postcode{DK-2300}
}

\author{Guido Salimbeni}
\email{guido.salimbeni@nottingham.ac.uk}
\orcid{0000-0001-6666-3518}
\affiliation{%
  \institution{University of Nottingham}
  \streetaddress{}
  \city{}
  \country{United Kingdom}
  \postcode{}
}

\author{Anders Sundnes Løvlie}
\email{asun@itu.dk}
\orcid{0000-0003-0484-4668}
\affiliation{%
  \institution{IT University of Copenhagen}
  \streetaddress{Rued Langgaards Vej 7}
  \city{Copenhagen}
  \country{Denmark}
  \postcode{DK-2300}
}

\author{Steve Benford}
\email{steve.benford@nottingham.ac.uk}
\orcid{0000-0001-8041-2520}
\affiliation{%
  \institution{University of Nottingham}
  \streetaddress{}
  \city{}
  \country{United Kingdom}
  \postcode{}
}

\author{Jichen Zhu}
\email{jicz@itu.dk}
\orcid{0000-0001-6740-4550}
\affiliation{%
  \institution{IT University of Copenhagen}
  \streetaddress{Rued Langgaards Vej 7}
  \city{Copenhagen}
  \country{Denmark}
  \postcode{DK-2300}
}

\renewcommand{\shortauthors}{Sivertsen et al.}

\begin{abstract}
Ongoing efforts to turn Machine Learning (ML) into a design material have encountered limited success. 
This paper examines the burgeoning area of AI art to understand how artists incorporate ML in their creative work. Drawing upon related HCI theories, we investigate how artists create ambiguity by analyzing nine AI artworks that use computer vision and image synthesis.
\Revision{Our analysis shows that, in addition to the established types of ambiguity, artists worked closely with the ML process (dataset curation, model training, and application) and developed various techniques to evoke the {\em ambiguity of processes}. Our finding indicates that the current conceptualization of ML as a design material needs to reframe the ML process as design elements, instead of technical details. Finally, this paper offers reflections on commonly held assumptions in HCI about ML uncertainty, dependability, and explainability, and advocates to supplement the artifact-centered design perspective of ML with a process-centered one.}

\end{abstract}


\maketitle


\section{Introduction}

To meet the demands raised by new Machine Learning (ML) products, the Human-Computer Interaction (HCI) community is making ongoing efforts to turn ML into a {\em design material}~\cite{holmquist2017intelligence}.
Despite the intense interest, a growing body of research shows that ML is uniquely difficult to design with\cite{Dove2017,yang_ml_hard,benjamin2021machine,kuniavsky2017designing}. In a recent study with industry UX designers, researchers identified two key difficulties of ML as a design material: {\em capability uncertainty} (uncertainties surrounding what the system can do and how well it performs) and {\em output complexity} (complexity of the outputs that the system might generate, e.g., in adaptive systems)\cite{yang_ml_hard}. Due to these ML-specific difficulties, designers face obstacles in all design phases, from conceptualization to prototyping. \Revision{There is thus a need in the HCI design community for new ways of conceptualizing ML and design criteria for ML experiences.}

  
In this paper, we follow the precedents in interaction design and seek inspiration from art. HCI researchers have found that engaging art and art history can open up new generative ideas for HCI theory and practice (e.g., ~\cite{Gaver2003,bardzell_humanistic_2015,benford_performing_2011}), especially in domains traditionally dominated by discourses of engineering and productivity (e.g., digital fabrication\cite{devendorf2015reimagining,song2021unmaking}, electronics\cite{kang2022electronicists} such as Machine Learning\cite{audry2021art,caramiaux2022explorers,scurto2021prototyping}). 

This paper turns to ``AI art,'' an emerging umbrella term that describes the variety of artistic practices that use AI, including ML, to create aesthetic experiences\cite{zylinska_ai_2020,boden}. Among AI art, we focus on a particularly active area: {\em visual} artworks that are built on {\em computer vision} --- the technology that allows computers to make sense of images --- and {\em image synthesis} --- the technology that allows computers to create images from some form of user input. Following recent breakthroughs in high-quality image synthesis, there is a surge of AI art experiments with this technology\cite{bogost_ai-art_2019,cetinic_creating_art_ai,zylinska_ai_2020,oppenlander,audry2021art}. For brevity, we will refer to the two types of artworks as AI art hereafter unless specified otherwise. 

Our particular focus is to investigate 
\Revision{{\em how digital artists work with ML as their material to create ambiguity, a common quality of art}}. In this paper, we adopt Gaver and colleagues' description of {\bf {\em ambiguity}} as ``a property of the interpretative relationship between people and artefacts ...'' that is ``evocative rather than didactic, and mysterious rather than explicit''\cite{Gaver2003}. Simply put, artifacts using ambiguity support multiple interpretations by users. Previous HCI research has proved ambiguity and multiple interpretations to be a fruitful alternative to the usability principles in traditional UX design~\cite{Gaver2003,staying_open,aoki_ambiguity,boehner_ambiguity,jorge_ambiguity,ambiguity_and_proto-practices,diffraction-in-action,introceptive}. \Revision{Similarly, our investigation of AI art indicates alternative approaches to some of the widely accepted desiderata of ML in HCI, such as dependability and explanability.}

Specifically, this paper reviews nine AI artworks whose process has been clearly documented. We use Gaver and colleagues' theory of ambiguity\cite{Gaver2003} as a theoretical guide to select AI artworks that cover a wide variety of tactics to create ambiguity. \Revision{Using the humanistic tradition of textual analysis and art critique, we study these artworks to understand how the artists make specific choices during the process of making the artworks to create ambiguities in the way the artworks can be understood. We identify techniques of how these artists used the entire ML process --- including dataset curation, model training, and application --- to evoke ambiguity in novel and creative ways. We reflect on the wider implications of our findings for HCI and design beyond art.}

The main contributions of this paper are the following:
\begin{enumerate}

    

    
    \item \Revision{We reveal how these AI artists engage ML in their creative process through analysis of nine recent AI artworks. Our analysis shows that the artists work closely with the entire ML process and identifies a list of techniques they have used to foster rich interpretations of AI by audiences. It demonstrates how ambiguity can be engaged with and materialized when dealing with ML in artistic and creative practices.
 }

    \item \Revision{We extend established HCI theories by identifying {\em ambiguity of process} as a new type of ambiguity particularly relevant to ML experiences. It indicates that, as the  HCI community wrestles with how to better use ML as a design material, the artifact-centered understanding of ambiguity and interpretation needs to be supplemented by a process-centered perspective that accounts for the ML process and how it is designed.}

    \item \Revision{We challenge widely accepted desiderata of ML in HCI, such as dependability and explanability, inspired by the alternative approaches demonstrated by the AI artworks we analyzed. }

\end{enumerate}

\section{Theoretical Framework}
We review three areas of related work: ambiguity as a resource for design in (primarily non-AI) HCI; accounts of how artists are evoking ambiguity within AI art; and uncertainty in the ML process.

\subsection{Ambiguity in HCI} \label{ambiguity_Lit}

Ambiguity is recognized as a valuable resource in HCI and design research\cite{Gaver2003,staying_open,aoki_ambiguity,boehner_ambiguity,jorge_ambiguity,ambiguity_and_proto-practices,diffraction-in-action,introceptive}. 
The notion that ambiguity can be a valuable resource for design, other than being a problem to be solved, was proposed by Gaver and colleagues in 2003\cite{Gaver2003}. 

Drawing inspiration from 
how artists employ various forms of ambiguity to make their artworks open to multiple interpretations by viewers, the 2003 paper identified tactics for using ambiguity as a resource for design. The paper argued for considering three broad types of ambiguity when designing interactive systems: 1) {\em ambiguity of information}, in which information may be portrayed deliberately blurry or, conversely, overly precise ways (sometimes appearing to be too precise can be ambiguous); 2) {\em ambiguity of context}, in which interactive artifacts are experienced in unusual contexts, including ones for which they were not originally designed; and 3) {\em ambiguity of relationship}, in which the user's relationship to the work is ambiguously framed. The paper also articulated a suite of tactics for deliberately employing ambiguity, which is listed in the appendix.

This framework has since been widely adopted in the HCI community. Subsequent work by Sengers and Gaver \cite{staying_open} explored how ambiguity can provoke interpretation. Their main idea is that an interactive experience can ask questions and thus prompt users to try and establish answers rather than aiming to give answers directly. Later research in HCI has further explored ambiguity from a variety of perspectives \cite{aoki_ambiguity,boehner_ambiguity,jorge_ambiguity}, including the areas of affective computing \cite{ambiguity_and_proto-practices}, bio-data \cite{diffraction-in-action,introceptive} and  experience design for museums \cite{sensitive_pictures,ryding_interpersonalizing_2021,ingimundardottir_word_2018,vayanou_storytelling,jorge_ambiguity}. 

\Revision{
With a few exceptions, Gaver and colleagues' framework has not been applied to AI or challenged in substantial ways. 
One exception built on Gaver at al’s original framework to reveal how creative dialogues between generative AI and humans, alongside the distinctive materiality of watercolour as a medium, generated drawings that exhibited multiple forms of ambiguity ~\cite{dyer2021signing}. Of particular significance here, one recent critique of Gaver at al’s orginal ambiguity paper highlights the importance of ‘prioritising the role of ambiguity in the process of designing’ rather than solely focusing on the ambiguities inherent in the designed artefact that emerges from these processes ~\cite{ drawing_conversations_ai}.
By applying Gaver et al's framework to the new context of AI art, our paper recognizes this lack of consideration of process. Our analysis of AI artworks demonstrates that, in addition to the three above-mentioned types of ambiguity, artists regularly use the ML process to evoke ambiguity. Our paper thus extends \cite{Gaver2003} by proposing a new type of ambiguity ----  ambiguity of process --- for artists and designers who wish to continue tapping into ambiguity as a design resource for ML experiences.} 


\subsection{Ambiguity in AI Art Theories}\label{sect:amb_artTheory}

In recent years, the field of AI art has seen a remarkable surge in innovation and public interest. 
The AI art landscape is enriched by artworks such as ``The Next Rembrandt,'' a painting generated using data from Rembrandt's entire body of work \cite{narvaez2022painting}, and ``Sunspring,'' a short film scripted by an AI \cite{brynjolfsson2017artificial}. 
Extensive reviews already exist on AI/ML art in general~\cite{audry2021art,cetinic_creating_art_ai}, artificial life/genetic art~\cite{tenhaaf2008art,penny2009art}, robotic art~\cite{penny2013art}, and artists' account on how they work with ML~\cite{caramiaux2022explorers}. 
It is important to note that we do not claim that ambiguity in AI art is categorically different from closely related art forms such as software art, evolutionary art, or glitch art. Many experiments in AI art continue the conceptual themes and approaches that these art movements have explored, and share similar intellectual traditions. 

While theories of AI art are still emerging, ambiguity has appeared as a central characteristic of aesthetic ML-based experiences. Our analysis of existing literature shows that current AI art theories on ambiguity can be broadly divided into three themes: 1) ambiguity as a defining aesthetic of the {\em ML output}, 2) ambiguity in the art discourse, and 3) ambiguity in the {\em ML process}.

First, artists and researchers have looked to ambiguity in the ML {\em output} as a defining aesthetic of AI art. For instance, Hertzmann \cite{hertzmann_visual_2020} notes that since modern ML does not have any concept of ``objects'' or ``space'', it can generate images that defy coherent spatial interpretation in ways similar to a painting by Escher. 
Hertzmann suggests this type of visual ambiguity should be championed as an inherent aesthetic of Generative Adversarial Networks (GANs). Similarly, Mazzone \cite{mazzone2019art} proposes that AI artists should embrace {\em stylistic ambiguity}, a strategy for ML to generate images based on blending recognizable artistic styles. This way, AI art can achieve novelty without departing too much from acceptable aesthetic standards. 
Furthermore, modern ML enables artists to process complex human languages, which contain many semantic ambiguities, in unprecedented ways. For example, Xu \cite{xu2021exploratory} intentionally used ambiguous sentences and unusual word combos as text prompts to generate images using {\em DALLE-2}. Murray-Browne and Tigas \cite{murray2021emergent} suggest that embracing ambiguity is valuable in computer vision-based interactive art installations involving body movement. They point to design for emergence, openness, and ambiguity to create opportunities for forms to emerge that go beyond the initial artistic vision.

Second, artists leverage the broader social discourses around AI/ML to frame audience's perception of AI art as ambiguous. 
For instance, Stark \cite{stark2019work} surveys how artists investigate the ethical ambiguity of data and ML as the intent of their work. 
Another common approach is to explore the ontological ambiguity around AI and AI art through the lens of anthropomorphism, especially the audience's tendency to attribute general human intelligence to existing narrow AI \cite{grba2021information,grba_ambiguities_creative_ai,grba_ai_art}. 
Continuing the practices of early AI artists such as Harold Cohen \cite{Cohen1995}, some contemporary artists deliberately withhold or obfuscate the information about how ML produces the artwork to inject ambiguity in the public's interpretation of the work \cite{cook2019framing}. Similarly, a recent study \cite{gu2022made} investigates how the audience's appreciation of a painting is affected when the viewer is uncertain about whether an AI or a human produced the artwork. 
Yurman and Reddy \cite{drawing_conversations_ai} explore using GAN-generated images as ''more than human'' elements to mediate the drawing conversations between two humans. In this work, the deliberate inability of the AI to produce clear and coherent images adds to the ``interpretative flexibility'' of the images sent between humans. 

Finally, the ML computational process has also been explored to evoke ambiguity. For example, Boyé and colleagues \cite{boye2019machine} investigate how artists explore machine flaws, irregularities, and errors in the computational process to push the boundaries of their artistic practice. Continuing the conceptual tradition of earlier digital art movements such as glitch art, ML's calculation errors become a creative opportunity to find new thematic, technological, and conceptual foundations for experimentation. \Revision{This paper extends this line of research by offering a close analysis of nine AI artworks to identify how AI artists tap into the ML process to create multiple interpretations and meanings.} To our knowledge, this paper is among the first works to take a closer look at the ML process and illustrate how artists' choices during the process lead to ambiguity in AI art. 
%



\subsection{Uncertainty in the Machine Learning Process}\label{sect: ML pipeline}

\begin{figure}
    \centering
    \includegraphics[width=\columnwidth]{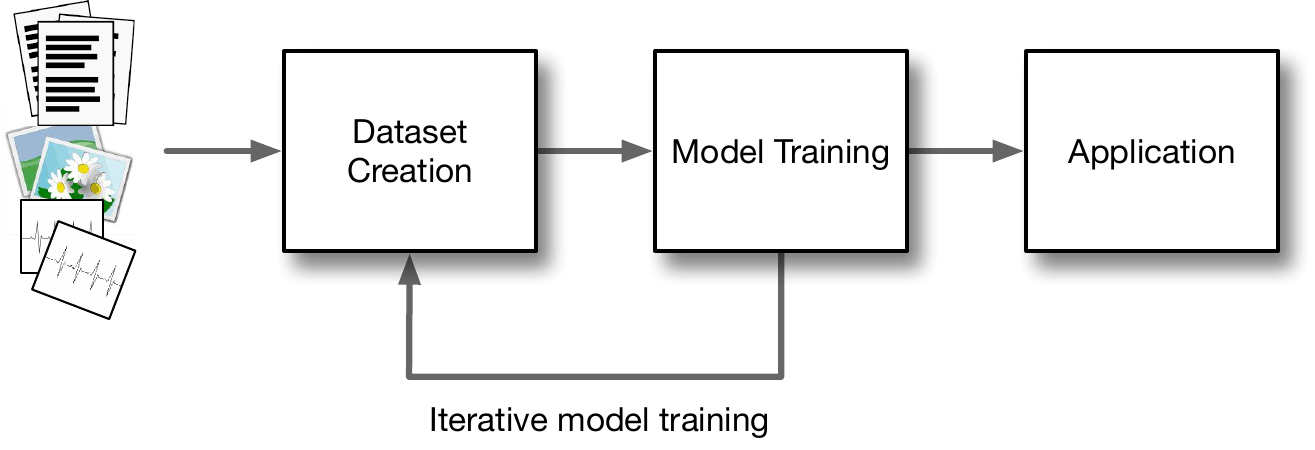}
    \caption{A Typical Machine Learning Pipeline for Computer Vision and Image Synthesis}
    \Description[An image describing the typical machine learning pipeline.]{The image shows the three main components of the deep learning pipeline, linking from 1) to 3): 1) dataset creation: collecting, cleaning, and labeling data to create a training dataset; 2) model training: select a model architecture and train a corresponding model using the training dataset; 3) application: apply the trained model for a specific task.}
    \label{fig:ML-pipeline}
\end{figure}

This section introduces the standard ML pipeline with key terminologies as well as the role of uncertainty in the ML process. In the rest of the paper, we use the term {\bf {\em uncertainty}} to refer to a property of the {\em ML/AI computational process}. In contrast, {\bf ambiguity} refers to the {\em interpretive relationship} between a) human audience and b) AI artworks \Revision{and the underlying ML process}. This is aligned with Gaver and colleagues' distinction of ambiguity from related concepts such as ``fuzziness'' or ``inconsistency,'' wherein they noted that ''ambiguity is an attribute of our interpretation'' whereas the latter concepts are attributes of things\cite{Gaver2003}.

Benjamin and Berger~\cite{benjamin2021machine} divide ML uncertainty into two main types. The first is {\em data uncertainty}, related to the ``noise'' in the training data sets. Data uncertainty can be introduced in the training dataset by, for instance, blurry images or erroneous labels. The second, and perhaps the more entrenched type, is what they call {\em model uncertainty}. It captures the ``epistemic uncertainty'' that characterizes all statistical inferences that underlie all ML decision-making. The ML process uses ``statistical intelligence''~\cite{Dove2017} to quantify, analyze, and manage uncertainty. How the engineers train a model, which ML architecture they choose to use, and when they stop the model training process can impact the amount of model uncertainty, but can never eliminate it completely. 
Benjamin and Berger argue that this type of statistical uncertainty is the ``material expression of ML decision making''\cite{benjamin2021machine}. 



The AI art in this paper uses Deep Neural Networks (DNN) models, also known as deep learning models. Fig. \ref{fig:ML-pipeline} shows the high-level technical process that all DNNs follow: 1) {\em dataset creation}: collecting, cleaning, and labeling data to create a training dataset; 2) {\em model training}: select a model architecture (e.g., GAN) and train a corresponding model using the training dataset; 3) {\em application}: apply the trained model for a specific task (e.g., plant recognition). Typically these steps are repeated iteratively until the model achieves the desired performance. In the following we will use ``dataset curation'' about the first step to highlight, how this can be part of a creative process involving the appropriation of existing datasets, remixing, and creating from the ground up.  

For example, in the case of a supervised image classification model, engineers first need to collect and clean a large collection of images the ML model will likely see. Each image needs to be labeled, by humans, with relevant tags (e.g., ``{\em violet}'' or ``{\em over-watered}''). Then, given a model architecture (e.g., Convolutional Neural Networks\cite{lecun1995convolutional}), engineers train the model iteratively using the training dataset to adjust the model's parameters to maximize its performance. 
After that, the model can be applied to recognize new images it has not seen before.  

\section{Approach} 
In order to investigate how AI artists use ML to create ambiguity, we selected nine artworks that use deep neural networks either for image synthesis or classification. Our first selection criteria is that the selected artworks should cover a wide variety of ambiguity. We used Gaver and colleagues' framework~\cite{Gaver2003} as a theoretical guide to select AI artworks that cover a wide variety of ambiguity types and tactics. Our second selection criteria is that we have access to the artist's technical description of how they used machine learning to create a specific artwork. This is to ensure that we can sufficiently analyze the underlying ML process. 

\Revision{In our selection process, the first, second, and third authors - who are HCI scholars with experience both as researchers and practitioners of AI art - used their domain knowledge to identify as many AI art projects as possible that match the second criteria above. Next, they individually analyzed each artwork and identified the evoked ambiguity types and which tactics the artists used \cite{Gaver2003} (details below). Next, all the authors collectively discussed their analyses and resolved disagreements until they reached an internal consensus.} Finally, we selected 9 artworks that altogether cover all of Gaver and colleagues' tactics except ``Point out things without explaining why'' (Tactic 9). While this tactic is not necessarily in conflict with our selection criteria stated above - that the artist must have published a technical description - we find in practice that these descriptions tend to also include explanations for the artistic choices in the work, to the extent that none of the selected artworks seem to match this tactic.
%

For analysis, we used \Revision{the humanistic method of textual interpretation and critique, which is commonly used for art analysis and increasingly adopted in HCI research\cite[19-20]{bardzell_humanistic_2015}. The methodological foundation for textual interpretation was originally laid by Gadamer in his theory of hermeneutics \cite{gadamer}. Hermeneutic interpretation has been used as a basis for humanistic HCI research in several influential studies \cite{staying_open, hook_horseback, pace_black_dress, mccarthy_wright} as well as in contemporary scholarship on AI Art \cite{zylinska_ai_2020,audry2021art,zhu2009intentional}. Specifically, we closely examined the artworks as well as their artists' technical descriptions of them. We then analyze 1) where in the ML pipeline the artists created the ambiguity we identified in the above-mentioned selection process, and 2) which technique they used. Similarly, the authors did independent analysis first and resolved any disagreements through discussion until concensus is reached.} A central debate in interpretation of art is whether or not interpretation should aim to identify the artists intentions \cite{lin_art_nodate}. In this paper we avoid drawing conclusions from the artwork itself to make statements about artistic intention - which has been criticised as the ``intentional fallacy'' - however we rely on statements made by the artists to understand how they have used the machine learning technology in the artistic process.
%
%
%
More details about the analysis can be found in the Appendix.  

\subsection{Sample AI Art}
Below we present the nine artworks. 

\begin{figure}
    \includegraphics[width=.6\columnwidth]{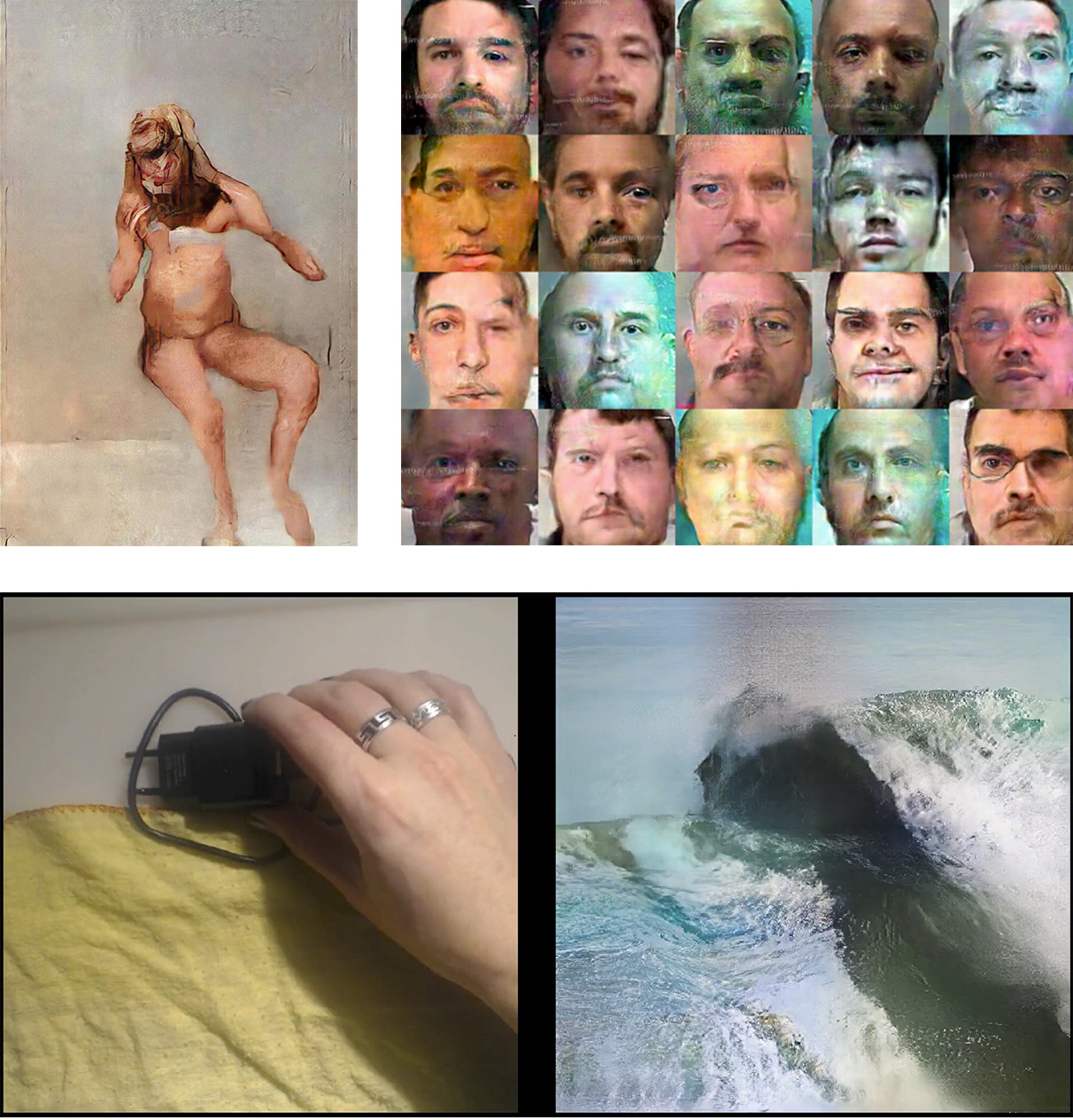}
    \caption{From top left to bottom: \textit{Butcher's Son} \cite{klingemann_butchers_2017}, \textit{Machine Bias} \cite{yaszdani_machine_2018} and \textit{Learning to See} \cite{akten_learning_2019}}
    \Description[A grid of images from each of the artworks Butcher's Son, Machine Bias, and Learning to See.]{In the top left is The Butcher's Son, a synthetically generated painting. Central in the image is a naked figure in a seated position. The face is obscured and parts of the torso and limbs are missing or transparent. The background is sand colored. The top right image shows a grid of distorted faces that have been generated by a GAN for the Machine Bias work. The bottom image is a still from Learning to See. It has two panels. On the left is a hand holding a mobile phone charger on a table. Below the hand is a yellow piece of fabric. On the right-hand panel, a texture looking like waves on a beach seems to mimic the shapes of the hand and charger.}
    \label{fig:first}
\end{figure}

\begin{figure}[h]
    \includegraphics[width=.6\columnwidth]{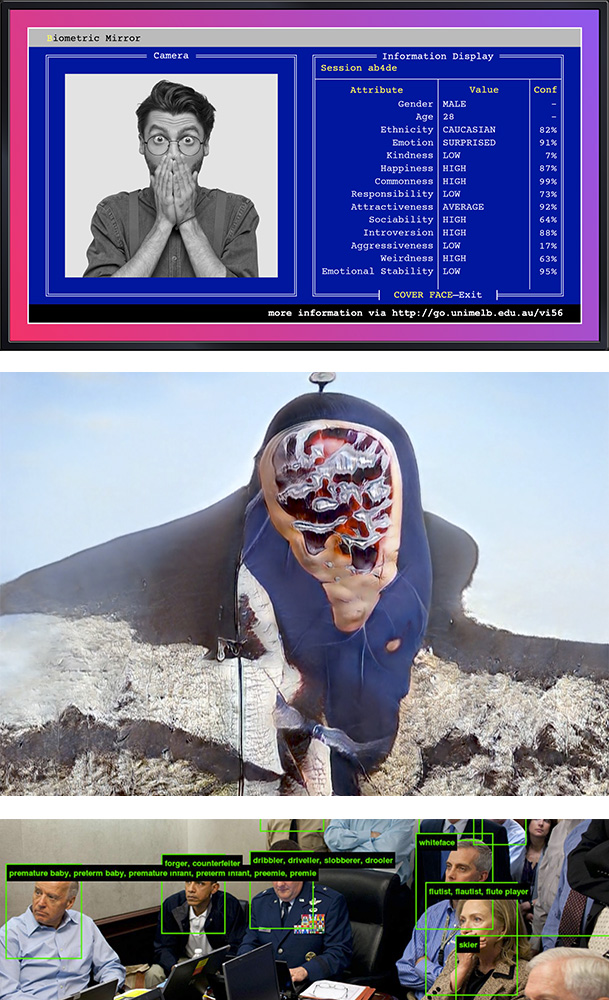}
    \caption{From top to bottom: \textit{Biometric Mirror} \cite{wouters_biometric_2019},\textit{POSTcard Landscapes from Lanzarote} \cite{guljajeva_postcard_2022} and \textit{ImageNet Roulette} \cite{paglen_imagenet_2019}}
    \Description[An image from each of the projects Biometric Mirror, POSTcard Landscapes from Lanzarote, and ImageNet Roulette arranged vertically]{ The top image is from Biometric Mirror. It shows a floating window with blue computer console output on a red and blue gradient. Inside the window on the left is a greyscale photograph of a man acting surprised. On the right-hand side is a list of labels and values. Some of the labels are ``Gender'', ``Emotion'', ``Attractiveness'', ``Introversion'', ``Weirdness'' and ``Emotional Stability''. Gender is shown as male, emotion as surprised, and the remainder is given a value of LOW, AVERAGE, or HIGH. The middle image shows an abstract figure in an image generated by a GAN from the POSTcards from Lanzarote project. The central figure has a blue body surrounding an area with a sand-colored rim and within that a reddish gradient with thick silvery lines. In the background are a rock-like texture and a wavy horizontal line. Above that is a blue sky. The bottom image is from ImageNet Roulette. This image shows a cropped version of the iconic Situation Room photograph from the White House on May 1, 2011. Several people in a small meeting room from the Obama administration are all looking at something to the left outside of the image. Each face is surrounded by a green bounding box with a label. Barack Obama is labeled ``forger, counterfeiter''. Joe Biden is labeled ``Premature baby, preterm baby, premature infant'' and similar terms. Hillary Clinton is labeled ``flutist, flautist, flute player''.}
    \label{fig:second}
\end{figure}

{\bf {\em The Butcher's Son}} (Fig. \ref{fig:first}, top left) is one in a series of generated painting-like portraits created by Mario Klingemann that ``focus on the human body, training his AI models to explore posture by turning stick figures into paintings, based on the analysis of images harvested from the internet'' \cite{klingemann_butchers_2017}. The images include visual artifacts introduced in the ML process.

{\bf {\em Learning to See}} (Fig. \ref{fig:first}, bottom) is an interactive video installation by Memo Akten \cite{akten_learning_2019} which appropriates GAN models trained on images of waves, flowers, and fire and applies them to real-time video feeds of mundane everyday objects such as phone chargers, pens, and fabric; turning them into animated waves, flowers, or fire in a similar composition.

{\bf {\em Machine Bias}} \cite{yaszdani_machine_2018} (Fig. \ref{fig:first}, top right) by Nushin Isabelle Yazdani is a series of generated faces based on photographs of prison inmates from across the US. She uses these ``future faces of prisoners'' to question predictive policing and automated pretrial risk assessment. 

{\bf {\em Biometric Mirror}} (Fig. \ref{fig:second}, top), created by Microsoft Research Centre for Social Natural User Interfaces \cite{wouters_biometric_2019}, presents itself as a system designed to ``stimulate individual reflection on the ethical application of artificial intelligence''. The system invites people to have their faces photographed through a webcam and analyzed by a psychometric system, which classifies their faces on a range of dimensions from relatively overt traits such as age, gender, and ethnicity to more diagnostic concepts such as aggressiveness, weirdness, and emotional instability.

{\bf {\em ImageNet Roulette}} (Fig. \ref{fig:second}, bottom) is a digital app and AI art installation by Crawford \& Paglen \cite{paglen_imagenet_2019}. The app lets users upload a photo, for instance, a selfie. The app will return the same photo with a green bounding box around every human face, each with a series of labels. For instance, in the iconic White House Situation Room photo Hillary Clinton is given the label ``flutist, flautist, flute player.'' Other users' photos received more problematic labels such as ``swot, grind, nerd, wonk, dweeb (...) rape suspect (...) first offender (...) gook, slant-eye'' \cite{wong_viral_2019}. These labels came from ImageNet, one of the most widely used {\em training datasets} in computer vision \cite{Crawford2019_excavating}.

\begin{figure}
   \includegraphics[width=.6\columnwidth]{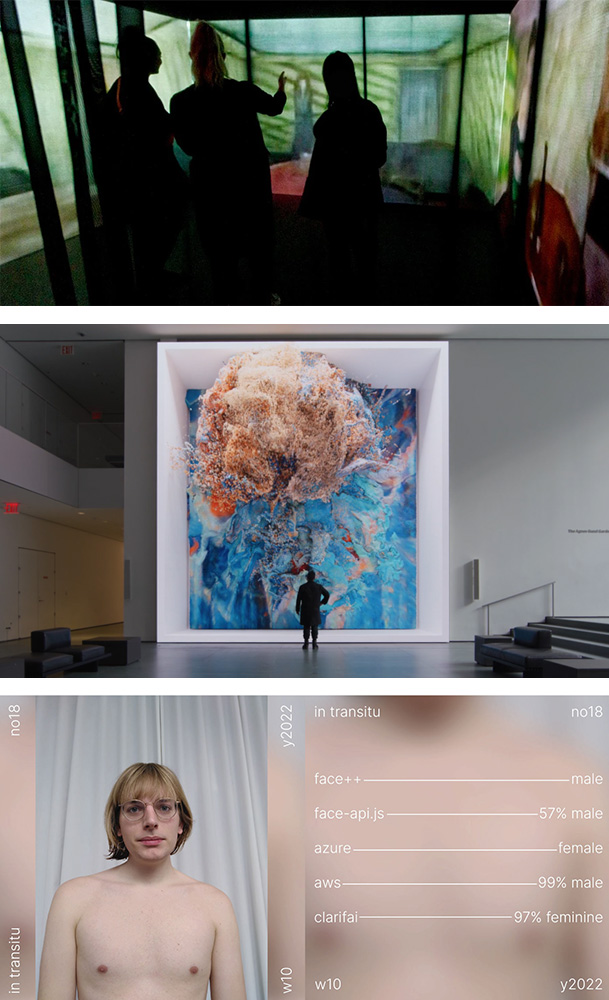}
    \caption{From top to bottom: \textit{Poison}\cite{munch_poison_2021}, \textit{Unsupervised}\cite{anadol_unsupervised_2022} and \textit{in transitu} \cite{ada_ada_ada_in_transitu_2022}}
    \Description[An image from each of the artworks Poison, Unsupervised and in transitu arranged vertically]{ 
    The top image is from Poison and shows the silhouettes of three people standing in a small room looking at projections of three paintings, one in front of them and one on each side. The projections are as large as the walls but look blurry. The middle image is from Unsupervised. It depicts a silhouette of a man standing in front of a large square digital screen in the lobby of MoMa. On the screen is shown an image of a white box that seems to extend backward from the surface of the screen, making a 3D effect. In the back of the box is a chaotic composition of colored particles mainly blue hues. At the top, a cloud of particles extends toward the viewer with red and sand hues. The bottom image shows a photograph of the nude torso of the artist Ada Ada Ada on the left hand. On the right-hand side are gender identification results from five different algorithms. Face++ labels the artist as male, face-api.js as 57\% male, Azure as female, AWS as 99\% female, and Clarifai as 97\% feminine. }
    \label{fig:third}
\end{figure}    

{\bf {\em POSTcard Landscapes from Lanzarote}} (Fig. \ref{fig:second}, middle) by Varvara \& Mar consists of two videos about the island of Lanzarote, part of the popular Canary Islands. It explores ``the tourist gaze'' in contrast to the local view of the place. The artists collected public photographs of the island on {\em Flickr} that represent these two perspectives \cite{guljajeva_postcard_2022}. The videos show the latent space of two generative models, each trained on one of the two sets of images. Each video shows morphing ambiguous imagery that at times vaguely resembles things like landscapes or airplanes, but often escapes clear categorization. 

{\bf {\em Poison}} (Fig. \ref{fig:third}, top) was a recent installation exhibited at the MUNCH museum in Norway \cite{sivertsen_art_2023}. It used ML to recreate Edvard Munch's ``Green Room'' paintings in a room in the museum. Visitors to the room saw projections of the ``Green Room'' from Munch's paintings covering three walls. As visitors moved through the room, the perspective of the artwork changed to match the visitor's movements, offering the illusion of stepping inside the room depicted in the paintings. In correspondence with the ambiguity of the paintings and questions raised in research about them, the perspective was unstable and the digital reproductions oscillated between different degrees of blurriness.

{\bf {\em Unsupervised}} \cite{anadol_unsupervised_2022} (Fig. \ref{fig:third}, middle) by Refik Anadol is - in the artist's terms - a ``data painting'' belonging to Anadol's series \textit{Machine Hallucinations}. Anadol has trained a StyleGAN 2 ADA model on 138,151 images from the archive of The Museum of Modern Art (MoMA). Through a custom piece of software called ``Latent Space Browser,'' the system generates images from the latent space of the model, resulting in fluid interpolations of colors, shapes, and patterns emerging from the corpus of the collection without ever representing any specific work as such.

{\bf {\em in transitu}} (Fig. \ref{fig:third}, bottom) by Ada Ada Ada \cite{ada_ada_ada_in_transitu_2022} explores gender recognition by commercial image analysis systems. The artist, a trans woman, periodically uploads images of herself naked from the waist up, along with the gender recognition outcome from five different ML models applied to the image. While the photos show only small variations in hairstyle and facial expression, the outputs from the algorithms vary widely. Sometimes the same photo is classified by different algorithms as male and female with confidence levels near 100\%. (All of the algorithms appear to treat gender as either a binary variable (male or female) or a percentage scale between male and female.) 

\section{Results}
\Revision{
Our selected AI artworks showcase how all three types of ambiguity - information, context and relationship - are invoked using the various tactics articulated by Gaver et al.\cite{Gaver2003} (for details see Table \ref{table:1} in Appendix).  
These artworks often evoke ambiguity of information. For example,  {\em Butcher's Son}, {\em Unsupervised}, and {\em POSTcard Landscapes from Lanzarote} did so through striking visual artifacts. We observed ambiguity of context in {\em Poison} and {\em Unsupervised} where models trained on paintings cast the art in a new role in the museum. 
Works such as {\em ImageNet Roulette}, {\em Biometric Mirror} and {\em Machine Bias} create ambiguity of relationship by placing the user/audience in an uneasy role as viewer and user of the artwork.  
}

When analyzing the artworks, it became clear that in our selected examples an important aspect of ambiguity is related to the machine learning processes. 
In this section, we will investigate how artists engage the computational process of data curation, model training, and application, to evoke ambiguity. 

\subsection{Dataset Curation}
The training datasets determine the ontology of an ML model --- what concepts and categories it can learn, what their relationships are, and what each concept ``looks like'' in the real world. In other words, datasets encode meanings, interpretations, and world views of those who made it~\cite{broussard2018artificial,Dignazio2020dataFem,crawford2021atlas,benjamin2019race,milan2019big}. Training datasets thus offer artists a rich source for ambiguity. 

\subsubsection{Selecting existing datasets with questionable ontologies} 
Training datasets establish the ground truth for ML models, but they are created by people in their social contexts. We find that some artists pick datasets to expose problematic value systems embedded within.  
A salient example is \textit{ImageNet Roulette}. The artwork used an ML model trained on a subset of {\em ImageNet} \cite{imagenet}, one of the most widely used {\em training datasets} in Computer Vision applications \cite{Crawford2019_excavating}. At the time of making the artwork {\em ImageNet} contained a vast 14 million labeled images, including a ``person'' category. The annotation of images with words has been done under the assumption that every verbal concept from the WordNet database could and should be ``imaged''\cite{maleve_data_2021}. The ``person'' category of labels included derogatory terms such as gendered and racial slurs which were applied to images scraped from the internet. This not only resulted in having images of specific people labeled with derogatory terms; it also made it possible for an algorithm trained on these categories to apply the same terms to images of other people. \textit{in transitu} and \textit{Biometric Mirror} similarly build on pre-existing models with problematic built-in assumptions: In the case of \textit{Biometric Mirror} the model relies on physiognomy -- the idea that a person's character can be assessed from their appearance; whereas the models explored in \textit{in transitu} rely on the assumption that gender identity is binary and unambiguous.    

Through the {\em hyper specificity} of the labels and scores in the datasets, a form of over-interpretation of the data exists at the very beginning of the ML process. 
By making visible the arbitrary and often offensive nature of the labels and scores these models assign to people, the artists encourage audiences to question the perception of computer vision as objective and neutral. In this way these three artworks expose the inconsistencies in the datasets' inherent world-view and cast doubt on their authority, utilizing tactics 2, 3 \& 4: \textit{Over-interpret data to encourage speculation}, \textit{Expose inconsistencies to create a space of interpretation} and \textit{Cast doubt on sources to provoke independent assessment}. 

\subsubsection{Making bespoke ontologies through new datasets} 
In our other examples, the artists took it upon themselves to collect, curate, and clean datasets that fit their purpose. However in \textit{Machine Bias} and \textit{POSTcard landscapes from Lanzarote} the artists make a point of creating datasets with an explicit ontological claim. In \textit{POSTcard landscapes from Lanzarote} the artists use photos from Flickr to build datasets that represent ``the tourist gaze'' and the local view, implying that the photos taken by visitors and locals reveal something about the ways in which these different groups of people see the place. However, a great part of the images are abstract and hard to interpret. As such, this use of data might be characterized as using tactic 1: \textit{Use imprecise representations to emphasise uncertainty}. 

\textit{Machine Bias} employs a similar strategy, building on a bespoke dataset - but arguably one which makes a much more problematic claim, suggesting that one may predict the facial features of future criminals based on analysis of the faces of past criminals. Thus the practice of collecting and applying this dataset raises philosophical and ethical questions about whether certain datasets should be collected and if it is even possible to represent with images what the data supposedly does. As they generate the faces of future criminals the artist intentionally \textit{over-interpret data to encourage speculation}(tactic 2).

Through these two techniques, the artists are investigating the ontology of their respective ML systems: what categories exist, how are they related, and what do they ``look'' like? Importantly, this investigation does not happen through descriptions of the formal qualities but by investigating what the models \textit{do}. How do they mediate the world they are supposedly representing?  

\subsection{Model Training}
Typically, an ML model is trained to optimize its performance defined by certain metrics (e.g., minimizing prediction error or creating a realistic-looking image). Several artworks engage with --- and often subvert --- the model training process in non-standard ways, often to create unusual or striking visual effects. 

\subsubsection{Repurposing upscaling to introduce visual artifacts} Upscaling is an ML technique to convert lower-resolution media to a higher resolution.
To save computational power, image synthesis models typically generate low-resolution images (e.g., 64x64 pixels) first and then use upscaling to enhance them to high-resolution ones (1024x1024 pixels) \cite{saharia_photorealistic_2022}. While upscaling is typically used to enhance images to more crisp and realistic looks, AI artists may repurpose the technique for different effects. Klingemann, the artist who created \textit{Butcher's Son}, used upscaling in a process he calls ``transhancement.'' As described in the artwork's catalog text, he used GAN to generate low-resolution content such as skin texture and hair and transhance them into a full portrait in ways that leave space for unusual artifacts: ``The result is painterly and ethereal, a neural network's vision of the human form.'' \footnote{https://www.artsy.net/artwork/mario-klingemann-imposture-series-the-butchers-son}. The resulting image carries an imprecise aesthetic and, from a human perspective, inconsistency in its seeming representation of a human body which are representative of tactic 1: \textit{Use imprecise representations to emphasize uncertainty} and  tactic 3: \textit{Expose inconsistencies to create a space of interpretation}.

\subsubsection{Under-fitting ML models} 
A common undesirable scenario in ML is \textit{over-fitting} a model. It happens when the model is tailored too exactly to the training dataset that it may fail to generalize to unseen data. Conversely, \textit{under-fitting} is another unwanted scenario where an ML model is unable to capture the relationship in the data accurately, thus having a low performance. It is usually caused by stopping the training process too early. 
\textit{Poison} turned this undesirable ML behavior into an artistic technique. The artists used it to emphasize the ambiguity of the motifs in Munch's paintings. The interactive visualizations were generated by applying style transfer to 3D renderings based on original paintings by Edvard Munch. However, instead of using the best-fitted model, the artists used an \textit{underfitted} version from earlier steps in the training process. This caused the resulting visualizations to have a fuzzy, ever-shifting appearance that never offered a clear representation of the motifs in the original paintings, thus applying tactic 1: \textit{Use imprecise representations to emphasize uncertainty}. 

\subsubsection{Changing the output modality from the input modality}
Both \textit{Poison} and \textit{Unsupervised} use static images of older artworks as their input (i.e. training data). However, in both works the output is dynamic and responds to the presence of visitors, thus including elements that are incongruous to the type of art they represent. The output of \textit{Unsupervised} is a dynamic visualization with mostly abstract, colorful patterns that are ever-emerging, moving, and disappearing. In \textit{Poison} the physical movement of the audience changes the perspective of the digital projections, mimicking the way one's perspective on a physical space shifts when moving around - offering an illusion of peering directly into the room depicted in the paintings. In both these cases, bespoke software was used to achieve these particular dynamics. Simultaneously, these visualizations refuse to reproduce the imagery of the original paintings, thus blocking them from being viewed in the way they were originally presented. Since the reproduced paintings in \textit{Poison} constantly shift and move, it removes the possibility of perceiving it the way you would perceive an oil painting. \textit{Unsupervised} removes the ability of the audience to recognize the artworks the system was trained on and presents them not as discrete objects, but as movements. From this perspective, \textit{Unsupervised} and \textit{Poison} might both be said to add and remove affordances to and from the originals they use as training data, leading to tactics 6 \& 7: \textit{Add incongruous functions to breach existing genres} and \textit{Block exposed functionality to comment on familiar products}. 

With these three techniques, the artists are exploring the concept of fit between the model and its underlying data. The fit is always characterized by some \textit{epistemic uncertainty} that can be reduced or increased. Here we see that contrary to conventional approaches a sub-optimal fit may be used for aesthetic effect.

\subsection{Application}
Through the selection of data used for an ML application, an ontology is established with an assumption that the model should be good at generating or classifying images within a specific domain. Shifting the domain can change the relationship in subtle or dramatic ways.

\subsubsection{Applying models to a different domain} 
Typically, ML engineers select the training datasets that resemble as much as possible the domain where the ML model will be deployed. However, AI artists have challenged this setup and applied models to different domains from where the training dataset was collected. An illustrating example is \textit{Learning to See}. The artist appropriates GAN models trained on images of waves, flowers, and fire and applies them to transform real-time video feeds of mundane objects into matching imagery of fantastic natural objects. The models consistently generate pretty visuals and effectively disguise the bland everyday objects. This technique may be seen as an expression of tactic 5: \textit{Implicating incompatible contexts to disrupt preconceptions}.  

\subsubsection{Connecting models directly to people}
Many surveillance applications quietly use ML to classify people, but those who were watched do not usually have access to how the ML model classifies them. Both \textit{in transitu} and \textit{Machine Bias} expose the functioning of ML systems by showing their direct application on concrete persons. Taking this technique one step further, \textit{ImageNet Roulette} and especially \textit{Biometric Mirror} encourage viewers to use \textit{their own} faces and observe how they will be classified. Having the rather stereotypical and sometimes even derogatory labels applied to their own body puts the audience in a vulnerable position that invites new perspectives on what impact such systems may have. These works may all be said to apply tactic 10: \textit{introduce disturbing side effects to question responsibility}.

When applying the system to an input found ``in the wild'', these two techniques show the opportunity to investigate what happens when we let the model mediate different parts of our world. New uncertainties can be generated through unconventional applications, and the promise of commercial systems can be scrutinized by applying them to situations that make the consequences of such mediation clear.

\subsection{From uncertainty to ambiguity}
Across all three steps of the process, the artists are negotiating the uncertainty inherent in the system by exposing it, exaggerating it, and generating new uncertainty by holding the model ontology against the world. They make some of these processes and considerations available to us, through the artwork itself, its staging, or metatext. This may, in turn, be experienced by us as ambiguous images and text (ambiguity of information), through systems that appear in unexpected contexts (ambiguity of context), and through the relation the system establishes with ourselves or other people (ambiguity of relation). By being exposed to these inconsistencies, unfamiliar relations, and imprecisions, we can doubt, question, speculate, and re-think our understanding of ML systems.   

\section{Discussion}
\subsection{Ambiguity of Process}
Our analysis shows ambiguity to be deeply embedded in and essential to the selected artworks. 
We found that Gaver and colleagues' framework remains a useful tool to understand the interpretative relationship between the audience and the artefacts. However, while ambiguity of information, context and relationship do explain to some extent how AI is interpreted by and through these various artworks, we contend that this is only a partial explanation.


Specifically, we propose that the much cited understanding of ambiguity set out in the original framework, and built upon by much subsequent work, is insufficient to fully capture how ambiguity is operating in these AI artworks. What is missing is a focus on process -- how artists creatively used the ML process to evoke ambiguity. Our analysis demonstrates how making AI is intimately bound up with complex technological processes, as is clearly evidenced by these various artworks. Moreover, and this is a key point, not only does ambiguity appear throughout the process, but the process is itself ambiguous, both to audiences and we argue to artists too.  The artists invoke ambiguities of the AI process itself to provoke audiences (and perhaps themselves) to interpret how AI creates artifacts. It is not just the artifacts that are ambiguous in various ways, but rather there is a deep and fundamental ambiguity about the AI process itself.

For example, works like {\em Machine Bias} rely on the process of the data curation being known, as the generated image itself without any context does not reveal the connections to predictive policing on its own. This is also the case for {\em Unsupervised} in which the choice of using data from the MoMA collection is important, and {\em Butcher’s Son} in which Klingemann’s use of transhancement plays an important role in the interpretations around the work. Similarly, the eerie visual artifacts in {\em Butcher’s Son} cannot be separated from how these images were generated and by whom. Techniques that we revealed above such as ``Selecting existing datasets with questionable ontologies'', ``Under-fitting ML models'' and ``applying models to a different domain'' directly create ambiguity concerning the ML process by which the artwork was generated.

There is of course a history of considering art as process rather than as artifact, notably in contemporary participatory and socially engaged arts \cite{birchall2017situating}. That this also applied to AI art has been discussed in Audry’s extensive survey of deep learning-based AI art\cite{audry2021art}. His work revealed artists’ active experimentation with almost every ML technical variation. Similarly, Caramiaux and Alaoui \cite{caramiaux2022explorers} documented the same trend from the artists’ perspective. They found that AI artists favor the process (i.e., the workflow) over the outcome as a way to create artistic experiences because it is difficult to anticipate the result of a specific model with a specific dataset. 

However, while recognized within the contemporary artworld, this focus on process has been largely absent from HCI’s consideration of ambiguity. Gaver et al.’s original ambiguity framework limited itself to considering how the artifacts produced by artists embodied different forms of ambiguity \cite{Gaver2003}. This, and subsequent work by Sengers and Gaver \cite{sengers2005reflective}, focused on how such ambiguity makes work open to interpretation by audiences. These lines of argument reflect a wider tendency within HCI to focus on the “meaning of the object produced” \cite{devendorf2015reimagining}. 

We therefore propose that HCI recognizes ambiguity of process as a fundamental and important additional type of ambiguity, alongside ambiguity of information, context and relationship. There are clearly complex relationships at play between these various types of ambiguity, evident in how our selected artworks invoke ambiguity of information, context and relationship, throughout what are also ambiguous processes. So we do not claim ambiguity of process to be a separate or orthogonal form of ambiguity. However, highlighting ambiguity of process inevitably shifts the focus of attention from designed artifacts to the processes by which they were made, while also extending the locus of interpretation to include the makers of artifacts who follow these processes (in our case artists, but perhaps more generally ``designers'') as well as their audiences (or perhaps more generally ``users''). 


\subsection{ML as a design material}

Recognizing the importance of the ML process, in addition to its output, has design implications for how HCI might engage with AI.
Despite the many benefits of conceptualizing ML as a design material \cite{Dove2017,yang_ml_hard}, this framing downplays the fact that ML itself is a technical process, an especially complex and opaque one --- including data curation, model training, and application. HCI literature tends to reflect the common practice where UX designers were often handed an already developed ML model with little access to how the ML model was trained. In this scenario, the design exploration is confined to tangible dimensions such as a system's input-output mapping (``possible system outputs'') and supported features (``system capability'') \cite{yang_ml_hard}. 

By contrast, the artists we analyzed showed that the ML technical process itself contains a multitude of design opportunities. In the AI technical community, researchers have been exploring AI as an expressive medium \cite{mateas2001expressive,vallgaarda2015interaction} and demonstrated that the AI technical process can profoundly impact its expressive affordance\cite{zhu2010towards}. However, this insight has not yet been widely adopted in HCI design. A process-centered conceptualization of ML as a design material can thus open up new design space by reclaiming the ML pipeline (data curation, model training, and application) as design elements. The techniques used by the AI artists to create ambiguity can be helpful to designers interested in evoking ambiguity and in critical design. 

How these artists directly engage various points in the ML process can inspire the broader HCI designer community. Decisions such as what kind of data should (not) be used in the training dataset, how much uncertainty and error is appropriate for the target users in context, and whether the model ontology aligns with the application domains are not just technical; they have a direct influence on the users and how they engage the system. It shows the need for designers to work more closely with the technical teams and gives concrete examples of some technical decisions that designers can engage from the vantage point of design, user needs, and ethics. 



\subsection{Uncertainty}
Our findings speak to a growing body of work that identifies uncertainty as an important property of ML as design material \cite{benjamin2021machine, caramiaux2022explorers,diffraction-in-action}. In contrast to existing literature that sees uncertainty primarily as a design challenge (e.g.,\cite{Dove2017,yang_ml_hard}), we argue that uncertainty is inherent to the ML process and cannot be fully eliminated. 
In many commercial applications of ML, this uncertainty is hidden. However, as Benjamin argues with the term \textit{horizontal relations}, our human-technology relations might be \textit{textured} with these uncertainties of ML systems as they are working behind our perceptual horizon. Common to many of the artworks we have examined here is the insistence on bringing back into view this texturing and exposing it through exaggerations and juxtapositions. The relation is brought back into our immediate, perceptual here and now through establishing other relations that bring the ML uncertainty back into view. As the uncertainty is exposed it gives rise to ambiguity. In this case, ambiguity is the most fitting way to represent the inherent uncertainty, rather than an incongruous insistence on accuracy and clarity. To the extent that we hide away the uncertainty in horizontal relations, we neglect our own ability as designers and as users to meaningfully assess the qualities of a given system.
Thus, while at first sight the concept uncertainty might seem to be strongly overlapping, even synonymous, with ambiguity, the former refers to sources of technical uncertainty within the AI technology, which  may in turn affect how certain or confident an AI system is of its own reasoning or outputs. In contrast, we treat ambiguity as a characteristic exhibited by a system that leads a human to make an interpretation of it. Such ambiguity may of course arise from technical uncertainties, but can also arise from many other sources (context, relationship, and now process).

\subsection{Dependability}

Like many other digital technologies before it, it is widely assumed that AI should be dependable, and there is much current concern about its tendency to make errors of judgement and other mistakes. When designing for probabilistic systems some degree of error is a given, and AI processes and interfaces may be designed to reflect this. For example, the typical ML training process involves careful, typically iterative, optimization to minimize errors based on some notion of ground truth, for example carefully curated and annotated training data. Dependability is of course an important requirement for many applications, such as safety critical systems or ones with other potentially damaging outcomes for people such as credit checking. However, and in stark contrast, errors, mistakes and frailties are grist to its mill for art where there is a rich history of creatively glitching digital technologies (to the point where ``glitch'' is a recognised genre of music). Our selected examples of AI art demonstrate that errors and shortcomings of the technology can help to define the aesthetics of ML art – whether it is to expose problematic aspects of the systems, as in \textit{ImageNet Roulette} or \textit{in transitu}, or exploiting the flaws or particular properties of the algorithms to create new aesthetic outputs, as in \textit{Learning to See} or \textit{Butcher’s Son}. This mirrors various proposals that AI should embrace failure. Hazzard and colleagues \cite{hazzard_failing} call for HCI to recognize the aesthetic potential of failure as a source of improvisation. Hertzmann \cite{hertzmann_visual_2020} suggests that improvements in the ability of image generators to create realistic images might force artists to start manipulating or ``breaking'' the algorithms to avoid producing images that look just like ordinary photographs, indicating that some current artists like Helena Sarin and Mario Klingemann are already exploring such approaches. Leahu \cite{leahu_ontological_2016} proposes that apparent failures in machine learning, such as the failure to reproduce one distinct object, can also be a consequence of a realist perspective, that is,
making the assumption that the world can be separated into discrete entities, and argue for a relational perspective that gives rise to ontological surprises as unexpected relations surface through the way ML systems make sense of their data. The ability of ML systems to convincingly produce outputs that are plausible imitations of things such as photographs calls for interfaces that embrace such relational perspectives, to work in tandem with the error-prone qualities of ML systems. 

Our selected artworks already reveal various ways of embracing error as a source of ambiguity. Looking beyond these, 
we highlight that instead of trying to hide ML errors away, designers can engage them purposefully to create various types of user experiences, especially in the context of home and play. For instance, previous ethnographic studies in HCI have employed conversation analysis to reveal how errors of virtual assistants such as Alexa become a source of humor when they ``fail'' during family dinner time conversations \cite{porcheron_voice_2018}. Game designers have engaged human players to detect ML errors as a form of playful experience\cite{villareale2022iNNk_MM,Zhu2021}. More broadly, researchers have found that exposing ML errors to users is helpful for them to develop accurate mental models of the system\cite{gero2020mental,villareale2022iNNk_MM}. Therefore, designers should use these AI artworks as a starting point to broaden the design space of how ML errors are exposed to users and how the ML system can recover from them.

\subsection{Explainability}
Another current focus of concern for human experience of AI arises from the black box and opaque nature of AI systems. Users, and indeed often developers, have little idea of how an AI produced a particular output. A widely adopted response is to apply eXplainable AI (XAI) techniques to open the black box by explaining ML models for human inspection \cite{Arrieta2020}. The perspectives adopted by our selected AI artworks suggest looking in the opposite direction, towards exploring the value of ambiguous and enigmatic AI that is not explained. Gaver, Sengers and others \cite{Gaver2003, staying_open} argued that ambiguity may inspire critical reflection and deeper engagement with systems. Similarly, one might consider how retaining AI’s unexplainable nature, or even amplifying it, might make AI more ambiguous and open to interpretation and so provoke humans to make their own interpretations, including questioning AI’s reliability and the nature of its biases. A growing body of literature suggests that users tend to over-rely on AI, which can be especially problematic in high-stakes areas \cite{robinette2016overtrust,howard2020we,sensitive_pictures}. Might it be that striving to make AI more explainable would exacerbate such problems, leading humans to believe even more in its ability to make decisions rather than trying to appreciate and accommodate its frailties and biases. Might ambiguously uncertain AI on the one hand or perhaps even overly confident AI on the other (thinking of Gaver et al’s tactic of sometimes being overly precise as a source of ambiguity) encourage users’ autonomy and critical thinking.
This line of argument mirrors thinking from Science and Technology Studies and media studies where researchers have long argued that AI and ML are discursive \cite{agre1997computation,Mateas2003a,zhu2009intentional} and that their proper functioning requires humans to interpret their algorithmic process to be ``intelligent'' and ``intentional.'' Recently, Murray-Browne and Tigas pointed out that key ML terms such as Training, Learning, Explanations, Bias and Black box are perhaps best considered as metaphors to help people interpret ML operations \cite{murray2022metaphors}.
More generally, we propose that there could be a greater focus on meaning making in relation to AI. For instance, in computer games the degree to which the user interface reveals the existence of the underlying AI can impact player-AI interaction \cite{Zhu2021}. For designers of creativity support tools, designing the appropriate interaction metaphor (e.g., nanny, pen pal, coach, and colleague \cite{lubart2005can}) can help users anticipate how to interact with the underlying generative ML models. Furthermore, we echo the point from Benjamin and colleagues \cite{benjamin_entoptic_2023} that metaphors are not just used to describe what AI technologies are but also what they do, thus reflecting the role of technology in actively shaping the world.

\section{Limitations}
This paper presents an analysis based on a fairly small collection of nine artworks. Even though this satisfies our goal of identifying techniques used by AI artists when working with ML, we acknowledge that a comprehensive analysis of a larger collection of artworks could reveal additional techniques to create ambiguity. Furthermore, we have focused mainly on art exhibited in recognized venues and/or created by established artists. This choice has left out the large, emerging field of amateur artists creating art with ML.

The exact steps that the artists took to make the artworks we analyzed are hidden from the public. Our analysis thus relies on the artists' public description of their work in artwork metatext, interviews, blog posts, academic publications - combined with our own technical knowledge. We acknowledge that there may be discrepancies in the artists' accounts and how the ML system actually functions. This is a methodological challenge since it is not feasible to verify whether the artists' descriptions are accurate. However, our analysis does not for the most part rely on fine details about the technical system, but on the main conceptual use of the technology in the artworks.   

\section{Conclusion}

We have revisited HCI's conception of ambiguity, bringing it to bear on machine learning, by examining the work of artists at the forefront of experimenting with this new technology. Through an analysis of nine AI artworks, we have identified multiple ways in which ambiguity emerges through the complex process of applying ML. This has led us to propose ambiguity of process as further fundamental form of ambiguity that is particularly relevant to AI because of it process oriented nature. In turn, recognising the inherent ambiguity of the AI process challenges contemporary thinking about AI, leading to the propositions that it might be improvisational rather than dependable, and interpretable rather than explainable.

We conclude by speculating that the ambiguous nature of AI that challenge both its dependability and explainability point to the fundamental existential question that underpins AI - that of ``intelligence''. There is a school of thought in contemporary AI that treats it as a tool or technology (or material as we considered earlier), an approach that naturally leans towards making AI more explainable, dependable, and certain (as we do with other tools). However the origins of AI lay in trying to make machines that in some ways mirror human intelligence. Artists are experts in exposing the human condition and they inherently recognise that human ‘thinking’ (as a human) is uncertain, unreliable and unexplainable. We often cannot explain ourselves (to ourselves let alone to others). We constantly seek to make new interpretations. We have more questions than answers. If AI is to be seen as being intelligent, then it needs to embrace the inherent and deep ambiguities of thinking.

\begin{acks}
We want to thank Peter Kun for his contributions in the early phases of the development of this paper. Furthermore, we appreciate the help from Tom Jenkins and anonymous reviewers for their valuable comments and criticism that have enabled us to sharpen the perspective and hone our argumentation.
\end{acks}

\bibliographystyle{ACM-Reference-Format}
\bibliography{bibliography}

\appendix
\section{Appendix}
\subsection{Method: Analysis of artworks}

\begin{table}[p]
 \caption{Overview of AI Art Examples Described in This Paper, in chronological order. "Types of Ambiguity" refers to the types identified in \cite{Gaver2003}. Similarly, the numbers under ``Tactics'' refer to the list of tactics in \cite{Gaver2003}, which is reproduced further down in this appendix.}
 \Description{The table shows a list of each analyzed artwork summarizing the year it was made, the technology used, the type of ambiguities and tactics we identified. Each row is a separate artwork.}
    \begin{tabular}{|  p{10em}  |p{2em}  |p{7em}  |l|p{10em}  |p{5em}  |} 
      \hline
      Work  &   Year    &    Technology      &     Types of Ambiguity  &   Tactics \\ 
      \hline
      \hline
      Butcher's Son\cite{klingemann_butchers_2017}      &   2017    &   Image synthesis     &   Information    &   1, 3   \\ 
      \hline
      Learning to See\cite{akten_learning_2019}      &    2017   &   Image synthesis     &   Information \& Context     &   1, 2, 5  \\ 
      \hline
      Machine Bias\cite{yaszdani_machine_2018}      &   2018    &   Image synthesis    &   Information \& Relationship   &  1, 2, 4, 10 \\ 
      \hline
      Biometric Mirror\cite{wouters_biometric_2019}     &   2018    &    Facial analysis    &   Information, Context \& Relationship   &   2, 3, 4, 5, 8, 10   \\ 
      \hline
      ImageNet Roulette\cite{paglen_imagenet_2019}      &   2019    &   Facial analysis    &    Information \& Relationship    &  2, 3, 4, 8, 10 \\ 
      \hline
       POSTcard Landscapes from Lanzarote\cite{guljajeva_postcard_2022}      &   2020    &    Image synthesis    &   Information  &   1, 2 \\ 
      \hline
      Poison\cite{sivertsen_art_2023}      &   2021    &   Image synthesis     &   Information \& Context   &   1, 3, 6, 7\\ 
      \hline
      Unsupervised\cite{anadol_unsupervised_2022}      &    2022   &    Image synthesis    &    Information \& Context   &   1, 6, 7\\ 
      \hline
       in transitu\cite{ada_ada_ada_in_transitu_2022}     &   2022    &   Facial analysis     &    Information \& Relationship     &   2, 3, 4, 10 \\ 
      \hline
    \end{tabular}
    \label{table:1}
\end{table}

The study reported in this paper had its outset in some reflections coming out of the authors' own design practices. The five authors of this paper come from backgrounds in HCI, design, computer science, media studies and art. In recent years we have in varying ways been exploring the qualities of AI as design material both as a theoretical research topic and in practice for applications relating to art, cultural heritage and games. Through informal, exploratory discussions we developed the initial theoretical focus of this paper: Examining what AI Art could teach us about AI as a design material, in light of the reflections on ambiguity in Gaver and colleauges' seminal paper from 2003.  

Using their domain expertise, the authors collectively identified a range of artworks that use ML in a broad variety of approaches. We included only artwork for which we could find some documentation or explanation of the use of ML, in order to ease analysis.

Using the list of tactics from Gaver and colleagues (see section \ref{sec:tactics} below) the authors chose a list of nine artworks which covered each of the ten tactics except no. 9: "Point out things without explaining why". While this tactic is not necessarily in conflict with our selection criteria stated above - that the artist must have published a technical description - we found in practice that these descriptions tended to also include explanations for the artistic choices in the work, to the extent that none of the selected artworks seemed to match this tactic. 

Next, these nine artworks were analyzed by three of the authors, of whom two have a background in art practice and the third is trained in humanistic media studies. For the most part the analysis was based on presentations of the artworks that could be found online, with the exception of \textit{Poison} which two of the authors could experience first-hand in the museum. Some of the artworks are fully available online: \textit{in\_transitu}, \textit{POSTcard Landscapes from Lanzarote}, \textit{Machine Bias}. For some of the other artworks we relied on images and video documentation of the artworks themselves (\textit{Butcher's Son}, \textit{Learning to See, Unsupervised}), as well as screenshots and documentation on the artist's websites (\textit{Imagenet Roulette}, \textit{Biometric Mirror}). Documentation of the ML process were sometimes presented in metatext accompanying the work, such as in the websites for \textit{Machine Bias}, \textit{Learning to See}, \textit{POSTcard Landscapes from Lanzarote}, \textit{Butcher's Son}, \textit{ImageNet Roulette}, \textit{in transitu}, \textit{Unsupervised} and \textit{Biometric Mirror}. Further explanations are frequently offered elsewhere in texts written by the artists such as essays \cite{Crawford2019_excavating,biometric_essay}, academic papers \cite{wouters_biometric_2019,learningtosee,guljajeva_postcard_2022,sivertsen_art_2023,sivertsen_iasdr_2023} and a PhD dissertation \cite{akten_phd}. 

For each artwork the three authors individually studied the artwork and the available documentation, using a hermeneutic approach as well as their prior knowledge about art and technology to interpret  the artwork and identify how ML was used to create ambiguity. Each of the authors independently identified which of the nine tactics this artistic practice could be said to employ, as well as which part of the ML process the artwork primarily highlighted.

Next, the three authors shared the results of their analysis with each other, and discussed the cases where their initial analysis diverged, until they reached consensus. 
    
Table \ref{table:1} provides an overview of the results.



\subsection{Ambiguity tactics}
\label{sec:tactics}

Below we reproduce the list of tactics provided by Gaver and colleagues \cite{Gaver2003}, along with the type of ambiguity they support: 
\begin{description}
\item [Information] 1) Use imprecise representations to emphasise uncertainty; 
\item [Information] 2) Over-interpret data to encourage speculation; 
\item [Information] 3) Expose inconsistencies to create a space of interpretation; 
\item [Information] 4) Cast doubt on sources to provoke independent assessment;
\item [Context] 5) Implicate incompatible contexts to disrupt preconceptions;
\item [Context] 6) Add incongruous functions to breach existing genres;
\item [Context] 7) Block exposed functionality to comment on familiar products;
\item [Relationship] 8) Offer unaccustomed roles to encourage imagination;
\item [Relationship] 9) Point out things without explaining why;
\item [Relationship] 10) Introduce disturbing side effects to question responsibility.
\end{description}

\end{document}